\documentclass{ws-procs975x65}

\usepackage{amsmath}
\usepackage{amsfonts}

\newcommand{\be}{\begin{equation}}
\newcommand{\ee}{\end{equation}}

\begin{document}

\title{POST-NEWTONIAN APPROXIMATIONS, COMPACT BINARIES, \\AND
STRONG-FIELD TESTS OF GRAVITY}

\author{Luc Blanchet} \address{Institut d'Astrophysique de Paris, 98
bis boulevard Arago, 75014 Paris, France\\ E-mail: blanchet@iap.fr}

\author{L P Grishchuk}
\address{School of Physics and Astronomy, Cardiff University, Cardiff CF243AA,
United Kingdom\\
{\it and} Sternberg Astronomical Institute, Moscow State University,
Moscow 119899, Russia\\
E-mail: grishchuk@astro.cf.ac.uk}

\author{Gerhard Sch\"afer}
\address{
Theoretisch-Physikalisches Institut,
Friedrich-Schiller-Universit\"at,
Jena, Germany\\
E-mail: Gerhard.Schaefer@uni-jena.de}

\begin{abstract}
This is an extended summary of the two
parallel sessions held at MG11: PPN1 ``Strong Gravity and Binaries''
(chaired by L.B. and L.G.) and PPN2 ``Post-Newtonian Dynamics in Binary 
Objects'' (chaired by G.S.). The aims and
contents of these sessions were close to each other and overlapping.
It is natural to review both sessions in one joint contribution to the MG11
Proceedings. The summary places the delivered talks in a 
broader perspective of current studies in this area. One can find more
details in individual contributions of the respective authors.
\end{abstract}

%\keywords{...}

\bodymatter

\section{Introduction and overview}
The current strong interest toward binary compact objects is largely
driven by the imminent observation of gravitational waves by the
presently operating or, more likely, by the upcoming advanced 
detectors. Many astronomical systems can emit some amounts of 
gravitational waves. However, it follows from
the most general theoretical considerations that the amount of
gravitational radiation is maximized when two massive chunks of matter
are moving with respect to each other with relativistic speeds. In
the Cosmos, this situation occurs naturally in tight binary systems
involving the most compact objects presently known --- neutron stars
and black holes. 

	This explains the special attention to the massive
compact objects orbiting each other at the late stages of their
relativistic evolution --- the so-called inspiralling compact
binaries~\cite{Clark79,Thorne87}. It is argued \cite{LPP97, GLPPS01}
that the inspiralling pairs of stellar mass \textit{black holes}
will probably be the first sources directly detected by the
ground-based laser interferometers. The theoretical description 
of such binaries is usually being based on the
Post-Newtonian (PN) approximations to general
relativity~\cite{Blanchet06,Damour87,Schaefer85}.
 
	In PN gravitational-wave studies it is often sufficient to treat
black hole as a highly compact massive object, independently of whether 
the object possesses the general-relativistic event horizon or not.
In the Newtonian limit, when formally $c\to\infty$, the Schwarzschild 
radius $r_g= 2Gm/c^2$ shrinks to zero, so the black hole can be 
viewed as a point-like particle endowed with a mass (and possibly 
a spin). The PN approximation is expected to provide a good description 
of black holes and their relative motion in an expansion series 
with $c\to\infty$, that is, when the dimensionless ratios,
such as $r_g/r$ and $v/c$, are sufficiently small.

	The masses of black holes are assumed to be ranging
from stellar masses (up to a few tens of M$_\odot$) to supermassive
black holes (SMBH) in nuclei of galaxies (around $10^5$ M$_\odot$ -
$10^9$ M$_\odot$). Although at present there is no decisive 
astrophysical evidence for the existence of compact objects 
with event horizons, the accurate PN equations of motion, as well as 
the emitted gravitational waveforms, are in principle capable of 
revealing the presence or absence of the event horizon. This is one 
of the features that makes these studies so exciting.

	The inspiralling compact binaries in the last minutes of
their orbital evolution are inherently powerful sources of gravitational 
waves. A crucial practical issue is the number of such systems 
that may populate a typical galaxy during a given interval of time, 
say, 1 year. Since the detectable part of gravitational radiation
from such sources is a relatively short-lived event, this number 
is called the event rate. If the event rate were too much low, 
the prospect of observing
these powerful sources would not be quite realistic. One would need to
survey a huge volume of the Universe in order to have some confidence
that at least several events occur per 1 year of observations. As a
consequence, the sensitivity of the instrument would be required to be
extremely high in order to guarantee the possibility of detecting the
sources located in the most remote parts of this volume.

	Unfortunately, the astrophysical event rates are not very 
certain ~\cite{NPS91, Phinney91, Kalogera04}. Sometimes their evaluation 
differs by orders of magnitude even in the papers of one and the same 
research group. This is especially true with regard to the binary 
systems involving black holes, for which the observational information is
scarce~\cite{BTKRB06}. However, it appears that there exists growing
consensus at least about the neutron star - neutron star (NS-NS)
coalescences. The currently most quoted rate $3 \times 10^{-5} -
10^{-4}$ per year, for a galaxy like our own Milky Way, is at the 
level estimated very early on ~\cite{Clark79, LPP87, HB90, TY93}
and advocated for long time by some groups~\cite{LPP97}.

	It also follows from general arguments based on binary star 
evolution and their numerical simulation that the black hole - black hole
(BH-BH) event rate is expected to be only an order of magnitude lower
than the NS-NS rate. If this is true, the larger total mass of the
BH-BH systems, in comparison with the NS-NS systems, and therefore
larger gravitational wave luminosity, can more than compensate their
lower event rate, when it comes to the analysis of detectability of
these systems. For an instrument of a given sensitivity, such as
LIGO~\cite{ligo} or VIRGO~\cite{virgo}, the probability of seeing a
BH-BH coalescence turns out to be higher than the probability of
seeing a NS-NS coalescence. Specifically, it was estimated that 
the BH-BH detection rate is likely to be a 
factor $\sim 5$ higher than the NS-NS detection rate, 
despite the opposite order of these systems event rates. 
This justifies the expectation that the first detected sources will
actually be the coalescing stellar mass black holes~\cite{GLPPS01}.
Theoretical and numerical studies of inspiralling and coalescing black 
holes are rising as a 
particularly interesting and important field of study. 

	Even the most powerful expected signals will be not more 
than at the level of noise in the presently available (ground-based)
instruments. To extract the signal from the noise, and to determine
the parameters of the radiating system, one needs to know in advance,
and as accurately as possible, the theoretical
signal templates for the incoming waves. It is also assumed that 
the response of the instrument to these waveforms is known with the 
equally high accuracy. The templates are being cross-correlated with the 
noisy output of the detector. Since the signal from inspiralling binaries 
is quasi-periodic, the knowledge of its phase is especially important,
and the templates must remain in phase with the expected true signal
as long as possible. The calculation of accurate templates from the 
compact binaries is now a matter of great activity.

	Processes involving supermassive black holes
(SMBH)~\cite{Rees07} provide more scientific oportunities, but add new
questions and complications. Gravitational radiation from a neutron
star or a stellar mass black hole inspiralling into a SMBH contains a 
wealth of information. A detailed waveform allows, in principle, to decide
whether the central object is a black hole or some even more exotic
object. Indeed, the multipole moments of a Kerr black hole satisfy
unique relationships as functions of its mass and spin, whereas the 
multipole moments of an arbitrary body are not, in general, linked by
similar relationships. The information about the multipole moments of the
central object is encoded in the waveform, thus providing an
example of strong gravity test with the help of gravitational
waves. In practice, one could start from evaluation of the quadrupole 
moment of the central object (which enters as a parameter in the 2PN 
waveform), and check whether it satisfies the conditions for a Kerr black 
hole. However, the complicated trajectories of infalling masses, the
uncertain radiation reaction force acting on the body, and the
inevitable presence of accreation disks surrounding real (in contrast
to ideal, theoretical) black holes, will make the extraction of
astrophysical information not so easy. 
           
	Though most of the current activities in relativistic gravity
are related to the imminent observation of gravitational waves, pulsar
astronomy also has the potential of further tests of gravitational 
theories and more advanced insights into the Einsteinian gravity. To 
mention in this respect are further observations of the double pulsar to 
eventually measure the moment of inertia of one of the pulsars which would 
give information on the equation of state of neutron stars~\cite{K06}, and 
the construction of SKA (square kilometre array) to detect practically 
all pulsars in our galaxy with a chance of discovery of pulsar-black-hole 
binaries~\cite{SKA}. 

\section{Theoretical modelling of compact binaries}
One normally makes a convenient, but approximate, separation
of the problem into several parts: internal, external, and far zone. 
This division is justified by the hierarchy of
characteristic length scales of the binary system: size of the bodies,
size of the orbit, and wavelength of the emitted gravitational
radiation. The internal part of the problem is concerned with the bodies 
themselves, including their sizes, shapes, internal structure and tidal
gravitational effects induced by one body on another. The external problem
deals with equations of motion of the \textit{centers of mass} of
participating bodies. As we said, in the external problem the bodies
are often treated as point particles with given masses. In advanced
treatments, the point particles are endowed also with spins and higher
multipole moments. Finally, the far zone problem is mostly
the calculation of the emitted gravitational waves. Typically this 
is being done in leading order in terms of the distance to the source. 

	Obviously, all three parts of the problem are interconnected. 
In particular, from the near-zone equations of
motion, or from the balance relationships equating the emitted
radiation and the changing orbital characteristics of the system, one
finds the radiation reaction force and corrections to the Keplerian
parameters of the binary, and then makes further corrections to the
waveforms. It should be qualified as a remarkable success that  
the problem of theoretical templates has been worked out by successive
approximations up to the 3.5PN approximation of general relativity,
corresponding to the accuracy $(v/c)^7$, where $v$ is the
relative speed of components of the binary. 

	The \textit{external} problem (that is, the motion of the centers 
of mass) has been solved at
3.5PN order independently by three groups, with completely equivalent
results. One group used the Arnowitt-Deser-Misner (ADM) Hamiltonian
formalism of general
relativity~\cite{JaraS98,JaraS99,DJSpoinc,DJSequiv} and worked in a
corresponding ADM-type coordinate system. Another group used a direct
PN iteration of the equations of motion in harmonic
coordinates~\cite{BFeom,ABF01,BI03CM}. Both groups used a
description by point particles and a self-field regularization. The
end results of these two approaches have been proved to be physically
equivalent~\cite{DJSequiv,ABF01}. However, both approaches left
undetermined one dimensionless parameter at the 3PN order. The
appearance of this unknown parameter was related with the choice of
the regularization method used to cure the self-field divergencies of
point particles. The completion of the equations of motion at
the 3PN order was made possible thanks to the powerful
\textit{dimensional} self-field regularization, which could fix up
uniquely the value of the ambiguity parameter in both
calculations~\cite{DJSdim,BDE04}. The third
approach~\cite{IFA01,itoh1,itoh2} succeeded in obtaining the equivalent
3PN equations of motion directly, \textit{i.e.} by using 
a ``surface integral'' method in
which the equations of motion are written in terms of integrals on
surfaces surrounding the compact bodies. This method is applicable
for extended compact objects in the strong-field point particle
limit. Finally, the 3.5PN terms, which constitute a 1PN relative
modification of the radiation reaction force (and are relatively
easier to derive), have been added in
Refs.~\cite{IW93,IW95,JaraS97,PW02,KFS03,NB05}.

The \textit{far zone} problem (that is, the
radiation field) has also been solved at the 3.5PN order,
\textit{i.e.} $(v/c)^7$ beyond the leading order given by the Einstein
quadrupole formula, using a particular gravitational wave generation
formalism combining multipolar expansions with the PN
approximation~\cite{B98mult}. The crucial step is the computation
of the binary's quadrupole moment at the 3PN order which has been done
by a combination of Hadamard's self-field regularization dealing with
most of the terms~\cite{BIJ02}, and eventually completed, like
in the problem of equations of motion, by dimensional regularization
able to fix the value of a few remaining ambiguity
coefficients~\cite{BDEI04}. The final 3.5PN templates~\cite{BFIJ02} take 
into account the values of the ambiguiy parameters computed 
in~\cite{DJSdim,BDE04,BDEI04} (for a review, see~\cite{Blanchet06}).

	However, there are also some difficult issues involved in 
this program. One should point out that the full analysis of the
\textit{internal} problem has
not been done at 3PN order and is hard to perform. It has been carried
out only for spherically-symmetric bodies and to the lowest radiative
order $(v/c)^5$, \textit{i.e.} to 2.5PN order in the equations of
motion. Satisfyingly, it was shown~\cite{GK83} that the compactness
parameter, characterising the size of the body $L$ in comparison with its
Schwarzschild radius $r_g$, can be absorbed into
the redefinition (or ``renormalization'') of the body's
mass. Therefore, at this level of accuracy, the equations of motion
are valid for compact objects of any size and structure, presumably
including black holes. This dependence of the results only on integral
parameters of the compact body (for example, its mass and spin) and
independence on its actual shape and internal structure (for example,
rearrangement of layers of matter) is sometimes called the principle
of ``effacement'' of the internal
structure~\cite{D83houches}. Obviously, the effacement principle
cannot be true with arbitrarily high accuracy. It has to be violated
at some sufficiently high order of PN approximations. 

	Despite the fact 
that the point-particle approach has reached a great level of rigor and 
sophistication, it cannot be arbitrarily accurate for
real physical objects, even if it is consistent mathematically for
idealized point particles. We have to worry about the magnitude of the 
finite-size effects, typically in the form of phenomenological parameters 
of an extended body and ultimately in the form of the finite gravitational 
radius of a black hole. Even though the PN corrections of very
high level of accuracy will not be required by observers in the near
future, further work is needed for precise identification of the
limits of theoretical consistency of the PN program itself. 
It is expected, though, that the trouble will not show up too early. 
Indeed, it follows from the simple arguments (see \textit{e.g.} 
\cite{GK83, Blanchet06}) that the tidal force in a pair of fluid
bodies is of the order of $\kappa(L/R)^5(GM^2/R^2)$, where $L$ is 
linear size of the bodies, $R$ - distance between them, and $\kappa$ is
a phenomenological parameter characterizing the ``elasticity" of the 
bodies. In the limit where $L$ approaches gravitational radius $r_g$, the 
tidal force is a factor $\kappa (v/c)^{10}$ smaller than the Newtonian 
force, \textit{i.e.} it is formally of the 5PN order and, hence, is
very small. This expectation is confirmed by explicit calculations of
finite-size effects for neutron star binaries \cite{MW04} and is
consistent with the relativistic equations of motion for black holes 
derived by matching of the perturbed Schwarzschild solutions \cite{D83houches}.

        It should be noted, however, that the above division of the 
problem into the internal and external parts certainly breaks down at 
the merger phase of compact objects. This is especially true for merging 
black holes and subsequent ringdown phase of the combined black hole.
This area of study requires new techniques and approaches 
(partially developed, see for example~\cite{DIS02, BCD06}). 
In particular, the merger and ringdown of binary black holes 
have recently been implemented by numerical 
techniques~\cite{Pretorius,Baker,Campanelli}. Although the amount of
gravitational radiation emitted at merger and ringdown phases 
is relatively modest, it may be observable by advanced detectors, 
opening an exciting era of new discoveries.

\section{Post-Newtonian equations of motion}
The introduction to the problem of testing general relativity with
gravitational radiation from compact binary inspirals was given by
\textit{C. Van Den Broek and B. S. Sathyaprakash} (reported 
by \textit{C. Van Den Broeck}). The speaker has emphasized 
that some of the PN terms in
the measurable gravitational wave phase arise from the scattering of
gravitational waves off the gravitational field of the source, in the
vicinity of the binary. These terms are known as
gravitational wave \textit{tails}. Observational checks of the
presence of such tail terms (using, for example, the data analysis method
of~\cite{Bsat95}) will be tests of the validity of PN approximations
and general relativity itself. The contribution of
\textit{C. Van Den Broek and B. S. Sathyaprakash} has also discussed
some other tests of gravitational theories, which are not necessarily 
based on PN approximations. The range of tested theories includes those
with massive gravitons. We will briefly review these possibilities below.

	\textit{S. Kopeikin} spoke about the irrelevancy of the
internal structure of gravitating bodies (\textit{i.e.} the effacement
principle) in the PN approximations of general relativity and
scalar-tensor theories of gravity. He argued that in general
relativity the effacing principle is violated by terms proportional to
the rotational moments of inertia of the fourth order. In the
scalar-tensor theories of gravity the violation begins earlier, by the
terms proportional to the second order rotational moments of
inertia~\cite{KV04}. When the effacement principle is violated, the
equations of motion of extended bodies differ from those of point-like
particles. Correspondingly, the emitted waveforms are also
different. In the limit where the size of the body is taken to be
close to the Schwarzschild radius $r_g = 2GM/c^2$,
\textit{Kopeikin} evaluates that the effacement principle is violated 
in the 3PN approximation in the case of scalar-tensor theories, and in
the 5PN approximation (terms of the order of $(v/c)^{10}$) in general
relativity. The latter statement is in agreement with an earlier 
conclusion by Damour~\cite{D83houches}.

	The point-particle approach inevitably encounters the
necessity of regularization of the fields diverging at the world lines
of the particles. There are two somewhat different methods of dealing
with this problem. One is directly applicable in the employed harmonic
coordinate system~\cite{BFeom,ABF01,BI03CM} while the other is
based on the ADM
formalism~\cite{JaraS98,JaraS99,DJSpoinc,DJSequiv}. Both methods
ultimately rely on analytical continuation of the equations of
motion to the (in general, complex) $d \neq 3$
dimensions~\cite{DJSdim,BDE04}. In his talk, \textit{P.~Jaranowski}
spoke about the dimensional regularization of the gravitational
interaction in the ADM formalism. More precisely he regularized the
3PN Hamiltonian of point masses, in which the field degrees
of freedom are reduced using the field equations~\cite{DJSdim}. The
speaker argued that the dimensional continuation leads to a finite and
unambiguous Hamiltonian in the limit $d \rightarrow 3$. He showed that
three somewhat different methods of computation lead to the same
expression for the dimensionally regularised 3PN Hamiltonian. This
increases confidence in the correctness of the 3PN Hamiltonian and the
associated equations of motion (an alternative method which confirmed
the result is~\cite{IFA01,itoh1,itoh2}).

       \textit{G. Faye} has derived the equations of motion of
spinning compact binaries including the spin-orbit (SO) coupling terms
1PN order beyond the leading-order effect~\cite{FBB06spin}. For black
holes maximally spinning this corresponds to 2.5PN order. The result
confirms the previous calculation of Ref.~\cite{TOO01}. The SO effects
up to 2.5PN order are also computed in the conserved (Noetherian)
integrals of motion, namely the energy, the total angular momentum,
the linear momentum and the center-of-mass integral. The spin
precession equations at 1PN order beyond the leading term are also
obtained. The speaker reported then the computation (using the
multipolar-PN wave generation formalism \cite{B98mult}) of 
the SO contributions in the gravitational-wave energy flux and the 
secular evolution of the binary's orbital phase up to 2.5PN
order~\cite{BBF06spin}. It was shown that the 1PN SO effects are in
general numerically larger than the spin-spin (SS) effects, in terms
of the number of gravitational-waves cycles, even though they appear
at a formally higher PN order. These results provide more accurate
gravitational-wave templates to be used in the data analysis of
rapidly rotating Kerr black-hole binaries with the ground-based
and space-based detectors.

	\textit{L. Gergely} discussed the corrections to the 2PN
equations of motion of structureless point-like particles, which arise
when the particles are endowed with spins and mass quadrupole
moments. If the body is a neutron star with a strong magnetic field,
then the representing ``particle'' is endowed also with a magnetic
type dipole moment. The speaker has demonstrated a generalized Kepler
equation~\cite{KMG05} where the orbital elements include contributions
from the spin-spin, mass quadrupole and magnetic dipole
interactions. The considered effects nicely fit in the standard Kepler
form with the modified parameters of the orbit.

	The orbital phase of inspiralling binaries with the inclusion
of the orbit's eccentricity was discussed in the contribution of
\textit{M. Vasuth}. He presented the results on the change of the mean
motion parameter for eccentric orbits and the evolution of the orbital
phase for circular orbits. All linear effects due to spins, mass
quadrupole and magnetic dipole moments are included in this
derivation. The author has related the change of the mean motion and
orbital frequency with the radiative energy losses. This allows one to
derive the time-dependent orbital frequency and phase, and to
calculate the number of relevant gravitational wave cycles. In the 2PN
approximation this number includes corrections from spin-orbit (SO),
spin-spin (SS), mass quadrupole - mass dipole and magnetic dipole -
magnetic dipole coupling effects. A special attention was paid to the
presence of a new self-interaction spin term~\cite{MVG05}.

	\textit{M. Tessmer} has reported on the accurate and
computationally efficient derivation of waveforms produced by binaries
with arbitrary eccentricity and mass ratio, and moving in slowly
precessing orbits~\cite{TG06}. The orbital motion is restricted to be
1PN accurate, and only the quadrupole contribution to the waves'
polarization amplitudes is taken into account. The central point of
the derivation is a special numerical method for solving the
participating generalized Kepler equation. The speaker has discussed
the relevance of the derived waveforms for observations with the
space-based interferometer LISA~\cite{lisa}.

	\textit{G. Sch\"afer} has reported on recent results from his
research group additionally to Tessmer's contribution, which were
the phasing of gravitational waves from inspiralling eccentric binaries
at the 1PN radiation-reaction order (corresponding to 3.5PN order
in the equations of motion)~\cite{KG06} and the gravitational recoil 
during binary black hole coalescence using the effective-one-body
approach~\cite{DG06}.

	\textit{L. Blanchet} discussed the problem of the gravitational
recoil of black-hole binaries using PN techniques. He reported on the
recent calculation of the recoil of non-spinning black holes in the
inspiralling phase at 2PN order\cite{BQW05}. The author estimated the 
kick velocity accumulated during the plunge from the innermost stable
circular orbit up to the horizon by integrating the momentum
flux along a plunge geodesic of the Schwarzschild metric. The
contribution to the total recoil due to the subsequent ringdown phase
was neglected.

\section{Supermassive black holes and strong field tests of gravity} 
In addition to possible SMBH mergers, a promising source for LISA,
called the extreme mass ratio inspiral (EMRI), is the infall of a
compact object into a SMBH. The infalling body can be a neutron star 
or a stellar mass black hole. The detailed study of gravitational waves 
emitted by an EMRI event can, in principle, allow one to build a 
``map'' of the gravitational field in a parsec radius of the
galactic nucleus. It is presumed that the SMBH candidate in the
nucleus is a Kerr black hole. However, the modelling of gravitational
wave from EMRIs is a very challenging task as the orbits often exhibit
complicated behaviour, and the EMRI parameter space is huge. In this 
situation, the derivation of true waveforms is practically impossible 
by analytical techniques and is very expensive by numerical methods.

	One approach to the problem is the construction of a family of
simpler, approximate waveforms, called ``kludge'' waveforms, which
nevertheless capture the main features of the true
signals~\cite{BFGGH06}. \textit{S. Babak} has discussed the progress
in compiling a bank of detection templates which are numerical
kludge (NK) waveforms. They can be generated quickly and cheaply, and
they are good enough to be used as a first pass test for the parameter
estimation. It is reported that satisfactory NK waveforms are
available for infalling masses in the range of up to 5-6 M$_\odot$.

	The derivation of sufficiently accurate waveforms always
requires the proper taking into account of the radiation reaction
force acting on the body and changing its trajectory. This problem 
is especially acute for EMRI sources where the inspiral
phase lasts for long time and orbits are complicated. In SMBH 
studies, one normally considers the technique of small perturbations 
of the background space-time. Due to the emission of gravitational 
waves, the world line of a compact object deviates in a calculable 
way from the nominal geodesic line of the background (possibly, a Kerr
black hole) solution~\cite{Mino05}.

	In the point-particle approximation, the metric perturbations are
divergent along the particle's world line. This requires some sort of
field regularization. Although many technical problems concerning the
regularization are already solved~\cite{MN98}, some conceptual issues
remain. \textit{Y. Mino} has reported on the existing problems for
the Kerr-background metric related to the choice of gauge, the restrictive 
assumption of linearity (weakness) of perturbations, and finally, the 
extraction of gravitational waveforms. One way to resolving these 
difficulties is a careful identification of the radiation-reaction 
part in the self-force expression. The author has also discussed the 
guides provided by the adiabatic approximation which assumes that the 
orbital parameters evolve slowly.

	We normally take it for granted that compact binary systems
are ``clean'', so that the relativistic gravity is essentially the
only participating interaction up to the very late stages of evolution
near the final merger. However, realistic astrophysical compact
objects are likely to be surrounded by accretion disks which will
complicate the motion of close companions and may intervene at the
level of corrections larger than the magnitude of relativistic PN
effects. This seems to be especially plausible to happen in the case
of EMRI sources, and this was the subject of the contribution by
\textit{P. Basu}. It is indeed expected that the late orbits of the
infalling object will be taking place within the accretion disk of the
SMBH. The infalling NS or BH will itself accrete some matter from the
disk. 

	Typically, the specific angular momentum of the accreted matter
is lower than the Keplerian angular momentum of the infalling body. 
Therefore, the total angular momentum of the infalling object will decrease 
more rapidly than expected, leading to a faster infall into the SMBH. For
some accretion disks the situation can be opposite, which would lead
to the slower than expected infall of the companion. \textit{Basu} has
outlined the gravitational field and fluid dynamics equations that
should be solved simultaneously. The gravitational field of the
central Kerr black hole is modelled by an effective Newtonian
potential~\cite{CM05}. For disks with certain parameters, the change
of the companion's angular momentum can be greater than the largest
losses expected due to gravitational radiation. Qualitatively, this
situation is similar to what can happen in neutron star systems when
interactions other than gravity are present~\cite{GLPPS01}.

\section{New research directions}
The overwhelming majority of current studies in gravitational physics
are based on the geometrical formulation of Einstein's equations. 
For example, in PN expansions describing an isolated binary system
(usually, in asymptotically Lorentzian harmonic coordinates), one
normally treats the gravitational functions $h^{\mu \nu}(t, {\bf
x})$ as frame-dependent pieces of curved space-time metric $g_{\mu
\nu}(t, {\bf x})$, rather than components of a genuine tensorial
gravitational field defined in a flat space-time with Minkowski
metric. 

	The thrifty geometrical picture of general relativity, which 
combines gravity, geometry and a choice of coordinates in a single 
mathematical object -- the curved space-time metric tensor 
$g_{\mu \nu}(x^{\alpha})$-- is accompanied by some peculiar features. 
One may recall ambiguities in the description of the gravitational field
energy density and gravitational-wave fluxes, constant mixing of
external (coordinate) and internal (``gauge'') transformations, and,
in general, certain disjointment of geometrical gravity from other
field theories.

	It is known for long time that the geometrical picture of
general relativity is by no means compulsory or necessary. It is
argued \cite{BG03} that the geometrical Einstein's gravity is fully 
equivalent mathematically and physically (including cosmology)
to a field theory in a flat space-time, \textit{i.e.} in a space-time
with zero-curvature metric tensor $\gamma^{\mu \nu}(x^{\alpha})$.
In flat space-time, one can always choose global Lorentzian coordinates, 
so that the metric tensor $\gamma^{\mu \nu}(x^{\alpha})$ takes on the 
familiar form of the Minkowski metric $\eta^{\mu \nu}$. The 
field-theoretical formulation of general relativity possesses all the 
necessary and strictly defined structures: covariant
gravitational Lagrangian containing gravitational field
$h^{\mu \nu}(x^{\alpha})$ and its first derivatives, second-order
differential non-linear field equations, gravitational energy-momentum 
tensor free of second and higher order derivatives of the field, 
universal coupling of the gravitational field to other physical fields
(which allows one to combine $h^{\mu \nu}(x^{\alpha})$ and
$\gamma^{\mu \nu}(x^{\alpha})$ into a single object $g^{\mu \nu}(x^{\alpha})$ 
and to reinterpret the theory as a geometrical theory,
where the curved space-time metric tensor $g_{\mu \nu}(x^{\alpha})$ 
satisfies geometrical Einstein equations), conservation laws, 
well defined and physically distinct coordinate and gauge 
freedoms. 

	One does not expect that any new observational conclusions 
will arise from a reformulation of one and the same fundamental 
theory, but a new angle
of view helps one to see problems in a different light and answer
questions which otherwise could not be even properly formulated.
In particular, the field-theoretical approach to gravity opens the
door for natural modifications of general relativity including the
concept of massive gravitons~\cite{BG03}. It is not surprising that 
what we are doing in practice, ``by hands'', in PN expansions is similar 
to traditional field-theoretical perturbative calculations, independently
of whether we adhere to geometrical ideology or not.

	The relevance of field-theoretical techniques for PN
calculations in binary systems was emphasized by
\textit{I. Rothstein} and \textit{R. Porto}. It was argued that the
replacement of detailed internal structure of the gravitating bodies
by a set of phenomenological parameters associated with a point-like
particle is similar to what is routinely being done in effective field
theories (EFT) where one is interested in a simple description of the
influence of short-scale physics on large-scale (low energy)
dynamics. The well developed methods of EFT allow one to introduce
simple power-counting arguments (evaluation of the order of magnitude
of various contributing terms) and employ traditional treatments of
divergencies and renormalization procedures~\cite{GR06,P06}. The
formalism also allows one to calculate absorptive effects for an
arbitrary object in terms of the graviton absorptive 
cross-section~\cite{GR06}. (In the context of gravity, the authors 
are using a somewhat confusing name of
``Non-Relativistic General Relativity''.) For application to PN
computations of compact binaries, these methods should be applied
twice. First, when one derives the equations of motion for point
particles, and second, when one calculates the gravitational waveforms
and replaces the entire binary system by a single particle with certain
multipole moments. The back-reaction effects within the EFT approch
should also be properly worked out.

	\textit{I. Rothstein} has stressed that the EFT calculations
are essentially being done at the level of the Lagrangian and the
action. He presented the action for a binary with slowly moving 
components. He argued
that the divergencies arising at the $(v/c)^6$ level, that is, in the
3PN approximation, are not physical as the corresponding terms in the
action can be removed by field redefinitions. This conclusion is
consistent with what was stated on the grounds of calculations in the
more usual PN approach~\cite{BFeom,BDE04}. Interestingly, 
the same line of EFT arguments has led the speaker to the conclusion 
that first terms which cannot be removed by field redefinitions, and
therefore provide a source of violation of the ``effacement"
principle, are of the order of $(v/c)^{10}$, \textit{i.e.} they appear 
in the 5PN approximation. Again, this conclusion is consistent with other
arguments, thus increasing our confidence in the internal workings of
the entire PN scheme.

	\textit{R. Porto} has extended the EFT approach to include
spin dynamics. This is achieved by adding rotational degrees of
freedom to the world-line action of the point-like particles. The
issue of different choices for the spin supplementary conditions was
clarified. It was shown that these conditions are equivalent, in the
sense of the final results, at least at the 1PN level. In the areas
where the EFT approach overlaps with previous
studies~\cite{KWW93,Kidder95,Pois98,WW07}, the final conclusions are
in agreement with each other. The author has also reported new
results, derived by EFT techniques, on the corrections to
the spin-spin (SS) potential in the 3PN approximation~\cite{PR06}. It
is argued that these corrections are easier to handle in the EFT
approach, and, in general, that this approach is a powerful tool to
treat in a systematic manner the higher-order PN effects.

	Spin effects encapsulated in a prescribed Lagrangian were also
considered in the contribution of \textit{M. Vasuth}. It is assumed
that the motion of the binary system is described by the
Lense-Thirring Lagrangian which takes into account the spin vector of
the central body and treats another body as a spinless test particle.
The author considers the 1.5PN accurate motion of the test particle
and concentrates on terms linear in the spin of the central body. The
radial and angular dynamics of the system are treated with the help of
more general results~\cite{GPV98}. The outcome of calculations are
both polarization components of the emitted gravitational waves,
including the contributions linear in the spin. It is confirmed that
the waveforms are in agreement with previous
calculations~\cite{Kidder95}.

	The geometrical general relativity requires one to be careful
with such things as choice of coordinates, identification of the
``gauge-independent'' degrees of freedom of the gravitational field,
clock synchronization, equivalence principle, and, in general,
physical interpretation of the participating quantities~\cite{L06}.
\textit{L. Lusanna} has discussed some of these issues in his
contribution. Although these issues refer in general to the full 
theory of geometrical gravitation, they also manifest themselves in 
approximations, such as PN gravitational-wave studies.

	Returning to the review of tests of general relativity, it is 
important to remember that interesting tests are not necessarily 
associated with the strong field regime. \textit{C.~Van Den Broek and
B. S. Sathyaprakash} have reminded us of the importance of some
signatures in the regime of weak plane gravitational waves. According to 
general relativity, the response of an interferometer to the incoming plane
wave contains not only the usual ``electric'' component but also the
(typically, smaller) ``magnetic'' component. The identification of the
``magnetic'' component in the output data is needed for proper
extraction of the radiating system's parameters, but also as a test of
the ``magnetic'' prediction of general relativity~\cite{BasG04}. 

	New possibilities arise in alternative theories of gravity, such 
as theories with massive gravitons. In particular, the detection of 
a ``scalar'' polarization state of gravitational waves, expected in
such theories, would revolutionize our views on the nature of gravity and 
could possibly provide a gravitational explanation to some presently existing
cosmological puzzles.

\section{Conclusions}
The current advances in observational facilities (LIGO and VIRGO on
Earth, LISA in Space) have stimulated further deep insights in such
problems as gravitational-wave physics in general and sources of
gravitational waves in particular, relativistic celestial mechanics, 
tests of general relativity and alternative theories. The contributions 
to the PPN1 and PPN2 parallel sessions have demonstrated the depth and 
rigor of the continuing research in these areas. Although several of the
recently derived results need to be cross-checked and placed in a
unique common context, it is clear that there is no major conceptual
or technical difficulties in this field. In particular the status of
PN computations of equations of motion and gravitational radiation
is quite satisfactory. In the coming years we will probably witness 
a major progress in relating and connecting the analytical PN calculations
with successful numerical computations. However, it appears that there 
is already enough theoretical clarity and completeness at
least for the level of accuracy of existing
gravitational wave observations. Hopefully, further theoretical work
will proceed hand in hand with successful experiments.


\begin{thebibliography}{99}

\bibitem{Clark79} J. P. A. Clark, E. P. J. van den Heuvel, and
W. Sutantyo, {\em Astron. \& Astrophys.} {\bf 72}, 120 (1979)

\bibitem{Thorne87}
K. S. Thorne,
in {\em Three hundred years of gravitation}, Eds. S. Hawking and W. Israel,
{\it Cambridge U. Press, 1987} p.330

\bibitem{LPP97}
V. M. Lipunov. K. A. Postnov, and M. E. Prokhorov,
{\em MNRAS} {\bf 288}, 245 (1997); {\em New Astronomy} {\bf 2}, 43 (1997)

\bibitem{GLPPS01}
L. P. Grishchuk, V. M. Lipunov. K. A. Postnov, M. E. Prokhorov, and
B. S.~Sathyaprakash, {\em Physics-Uspekhi} {\bf 44}, 1-51 (2001) 
(astro-ph/0008481) 

\bibitem{Blanchet06} L. Blanchet, in {\em Living Reviews in
Relativity} {\bf 9}, 4
(2006)\\(http://www.livingreviews.org/lrr-2006-4)

\bibitem{Damour87}
T. Damour,
in {\em Three hundred years of gravitation}, Eds. S. Hawking and W. Israel,
{\it Cambridge U. Press, 1987} p.128

\bibitem{Schaefer85}
G. Sch\"afer,
{\em Ann. Phys. (NY)} {\bf 161}, 81 (1985)

\bibitem{NPS91}
R. Narayan, T. Piran, and A. Shemi, {\em Astrophys. J.} {\bf 379}, L17 (1991)

\bibitem{Phinney91} 
E. S. Phinney, {\em Astrophys. J.} {\bf 380}, L17  (1991)

\bibitem{Kalogera04} V. Kalogera, C. Kim, D. R. Lorimer, {\em et al.},
{\em Astrophys. J.} {\bf 601}, L179 (2004) [ERRATUM: {\em Astrophys. J.}
{\bf 614}, L137 (2004)]

\bibitem{BTKRB06}
K. Belczynski, R. Tamm, V. Kalogera, F. Rasio, and T. Bulik,
arXiv: astro-ph/0612032

\bibitem{LPP87}
V. M. Lipunov. K. A. Postnov, and M. E. Prokhorov,
{\em Astron. \& Astrophys.} {\bf 176}, L1 (1987)

\bibitem{HB90}
D. Hils. P. Bender, and R. Webbink,
{\em Astrophys. J.} {\bf 360}, 75 (1990)

\bibitem{TY93}
A. V. Tutukov and L. R. Yungel'son,
{\em Astron. Rep.} {\bf 37}, 411 (1993)

\bibitem{ligo}
http://www.ligo.org, http://www.ligo.caltech.edu 

\bibitem{virgo}
http://www.virgo.infn.it

\bibitem{Rees07} M. J. Rees and M. Volonteri, ArXiv: astro-ph/0701512

\bibitem{K06} M. Kramer, I.H. Stairs, R.N. Manchester, M.A. McLaughlin,
A.G. Lyne, R.D. Ferdman, M. Burgay, D.R. Lorimer, A. Possenti,
N. D'Amico, J.M. Sarkissian, G.B. Hobbs, J.E. Reynolds, P.C.C. Freire,
and F. Camilo, {\em Science}, {\bf 314}, 97 (2006)

\bibitem{SKA} http://www.skatelescope.org/

\bibitem{JaraS98}
P.~Jaranowski and G.~Sch\"afer, {\em Phys. Rev. D}, {\bf 57}, 7274 (1998)

\bibitem{JaraS99} P.~Jaranowski and G.~Sch\"afer, {\em Phys. Rev. D},
  {\bf 60}, 124003 (1999)

\bibitem{DJSpoinc} T.~Damour, P.~Jaranowski, and G.~Sch\"afer, {\em
Phys. Rev. D}, {\bf 62}, 021501R (2000)

\bibitem{DJSequiv} T.~Damour, P.~Jaranowski, and G.~Sch\"afer, {\em
Phys. Rev. D}, {\bf 63}, 044021 (2001)

\bibitem{BFeom} L. Blanchet and G. Faye, {\em Phys. Rev. D}, {\bf 63},
    062005 (2001)

\bibitem{ABF01} V.C. de~Andrade, L.~Blanchet, and G.~Faye, {\em
Class. Quant. Grav.}, {\bf 18}, 753 (2001)

\bibitem{BI03CM} L. Blanchet and B. R. Iyer, \newblock {\em
Class. Quant. Grav.}, {\bf 20}, 755 (2003)

\bibitem{DJSdim} T.~Damour, P.~Jaranowski, and G.~Sch\"afer, {\em
Phys. Lett. B}, {\bf 513}, 147 (2001)

\bibitem{BDE04} L. Blanchet, T. Damour, and G. Esposito-Far{\`e}se,
{\em Phys. Rev. D}, {\bf 69}, 124007 (2004)

\bibitem{IFA01} Y.~Itoh, T.~Futamase, and H.~Asada, {\em
Phys. Rev. D}, {\bf 63}, 064038 (2001)

\bibitem{itoh1} Y. Itoh and T. Futamase, {\em Phys. Rev. D}, {\bf 68},
  121501R (2003)

\bibitem{itoh2} Y. Itoh, {\em Phys. Rev. D}, {\bf 69}, 064018 (2004)

\bibitem{IW93} B.R. Iyer and C.M. Will, {\em Phys. Rev. Lett.}, {\bf
  70}, 113 (1993)

\bibitem{IW95} B.R. Iyer and C.M. Will, {\em Phys. Rev. D}, {\bf 52},
  6882 (1995)

\bibitem{JaraS97} P. Jaranowski and G. Sch{\"a}fer, {\em
Phys. Rev. D}, {\bf 55}, 4712 (1997)

\bibitem{PW02} M. E. Pati and C. M. Will, {\em Phys. Rev. D}, {\bf 65}:104008,
  2002.

\bibitem{KFS03} C.~K{\"o}nigsd{\"o}rffer, G.~Faye, and G.~Sch{\"a}fer,
{\em Phys. Rev. D}, {\bf 68}, 044004 (2003)

\bibitem{NB05} S. Nissanke and L. Blanchet, {\em Class. Quant. Grav.},
  {\bf 22}, 1007 (2005)

\bibitem{B98mult} L. Blanchet, {\em Class. Quant. Grav.}, {\bf 15},
  1971 (1998)

\bibitem{BIJ02} L. Blanchet, B. R. Iyer, and B. Joguet, {\em
Phys. Rev. D}, {\bf 65}, 064005 (2002)

\bibitem{BDEI04} L. Blanchet, T. Damour, G. Esposito-Far{\`e}se, and
B. R. Iyer, {\em Phys. Rev. Lett.}, {\bf 93}, 091101 (2004)

\bibitem{MW04}
T. Mora and C. Will,
{\em Phys.Rev. D}{\bf 69} 104021 (2004)

\bibitem{BFIJ02} L. Blanchet, G. Faye, B. R. Iyer, and B. Joguet, {\em
Phys. Rev. D}{\bf 65}, 061501R (2002)

\bibitem{GK83} L. P. Grishchuk and S. M. Kopeikin, {\em
Sov. Astron. Lett.}, {\bf 9}(4), 230 (1983); in {\em Relativity in
Celestial Mechanics and Astrometry}, Eds. J. Kovalevsky and
V. A. Brumberg, {\it Reidel, 1986} p.19

\bibitem{Pretorius}
F. Pretorius, {\em Phys. Rev. Lett.}, {\bf 95}, 121101 (2005)

\bibitem{Baker} J. Baker, J. Centrella, D.-I. Choi, M. Koppitz, and
J. van Meter, {\em Phys. Rev. Lett.}, {\bf 96}, 111102 (2006)

\bibitem{Campanelli} M. Campanelli, C. O. Lousto, P. Marronetti, and
Y. Zlochower, {\em Phys. Rev. Lett.}, {\bf 96}, 111101 (2006)

\bibitem{D83houches} T.~Damour, in {\em Gravitational Radiation},
N.~Deruelle and T.~Piran, editors, p. 59, Amsterdam, North-Holland
Company (1983)

\bibitem{DIS02}
T. Damour, R. Iyer, and B. Sathyaprakash,
{\em Phys. Rev. D} {\bf 66}, 027502 (2002) 

\bibitem{BCD06}
A. Buonanno, Y. Chen, and T. Damour,
{\em Phys. Rev. D} {\bf 74}, 104005 (2006)

\bibitem{Bsat95} L. Blanchet, and B. S. Sathyaprakash, {\em
Phys. Rev. Lett.}, {\bf 74}, 1067 (1994)

\bibitem{KV04} S. Kopeikin and I. Vlasov, {\em Physics Reports}, {\bf
400}, 209 (2004); arXiv: gr-qc/0612017

\bibitem{FBB06spin}
G. Faye, L. Blanchet, and A. Buonanno, {\em Phys. Rev. D}, {\bf 74},
104033 (2006)

\bibitem{TOO01}
H.~Tagoshi, A.~Ohashi, and B.J. Owen, {\em Phys. Rev. D}, {\bf 63},
044006 (2001)

\bibitem{BBF06spin} L. Blanchet, A. Buonanno, and G. Faye, {\em
Phys. Rev. D}, {\bf 774}, 104033 (2006)

\bibitem{KMG05}
Z. Keresztes, B. Mikoczi, and L. Gergely,
{\em Phys. Rev. D} {\bf 72}, 104022 (2005)

\bibitem{MVG05}
B. Mikoczi, M. Vasuth, and L. Gergely,
{\em Phys. Rev. D} {\bf 71}, 124043 (2005)
 
\bibitem{TG06}
M. Tessmer and A. Gopakumar,
{\em MNRAS} {\bf 374}, 721 (2007)

\bibitem{KG06}
C. K\"onigsd\"orffer and A. Gopakumar,
{\em Phys. Rev. D} {\bf 73}, 124012 (2006)

\bibitem{DG06}
T. Damour and A. Gopakumar,
{\em Phys. Rev. D} {\bf 73}, 124006 (2006)

\bibitem{BQW05}
L. Blanchet, M.S. Qusailah, and C.M. Will, 
{\em Astrophys. J.} {\bf 635}, 508 (2005)

\bibitem{lisa}
http://www.lisa.jpl.nasa.gov

\bibitem{BFGGH06}
S. Babak, H. Fang, J. Gair, K. Glampedakis, and S. Hughes,
ArXiv: gr-qc/0607007

\bibitem{Mino05}
Y. Mino,
{\em Class. Quant. Grav.} {\bf 22}, S717 (2005)

\bibitem{MN98} Y. Mino, and H. Nakano, {\em Progr. Theor. Phys.} {\bf
  100}, 507 (1998)

\bibitem{CM05}
S. K. Chakrabarti and S. Mondal,
{\em MNRAS} {\bf 389}, 976 (2005)

\bibitem{BasG04}
D. Baskaran and L. P. Grishchuk,
{\em Class. Quant. Grav.} {\bf 21}, 4041 (2004)

\bibitem{BG03}
S. V. Babak and L. P. Grishchuk,
{\em Phys. Rev. D}{\bf 61}, 024038 (1999);
{\em Intern. J. Mod. Physics D} {\bf 12}(10), 1905 (2003)

\bibitem{GR06}
W. Goldberger and I. Rothstein,
{\em Phys. Rev. D} {\bf 73}, 104029 (2006); {\it ibid}, 104030

\bibitem{P06}
R. A. Porto,
{\em Phys. Rev. D} {\bf 73}, 104031 (2006)

\bibitem{KWW93}
L. Kidder, C. Will, and A. Wiseman,
{\em Phys. Rev. D} {\bf 47}, R4183 (1993)

\bibitem{Kidder95}
L. Kidder,
{\em Phys. Rev. D} {\bf 52}, 821 (1995)

\bibitem{Pois98}
E. Poisson,
{\em Phys. Rev. D}{\bf 57}, 5287 (1998)

\bibitem{WW07}
H. Wang and C. M. Will,
ArXiv: gr-qc/0701047
 
\bibitem{PR06}
R. A. Porto and I. Rothstein,
{\em Phys. Rev. Lett.}{\bf 97}, 021101 (2006)

\bibitem{GPV98}
L. A. Gergely, Z. Perjes, and M. Vasuth,
{\em Phys. Rev. D}{\bf 57}, 876 (1998)

\bibitem{L06}
L. Lusanna,
in {\em Current Mathematical Topics in Gravitation and Cosmology},
Karpacz Winter School, Poland, 2006 (gr-qc/0604120) 

\end{thebibliography}
\end{document}